\documentclass[referee,usenatbib]{mn2e}

\usepackage{graphicx}
\def\ls{{_<\atop^{\sim}}}
\def\gs{{_>\atop^{\sim}}}
\def\cgs{ ${\rm erg~cm}^{-2}~{\rm s}^{-1}$ }

\title[Rapid N$_H$ changes in NGC 4151]{Rapid N$_H$ changes in NGC 4151}

\author[S. Puccetti et al.]{S.~Puccetti$^{1,2}$, F.~Fiore$^2$, G.~Risaliti$^{3,4}$, M.~Capalbi$^1$,
\newauthor M.~Elvis$^4$, F.~Nicastro$^{2,4,5}$\\
$^1$ASI Science Data Center, via Galileo Galilei, 00044 Frascati Italy.
\\
$^2$INAF-Osservatorio Astronomico di Roma,via Frascati 33,
Monteporzio-Catone (RM), I00040 Italy.\\
$^3$INAF-Osservatorio Astrofisico di Arcetri, Largo Enrico Fermi 5,
Florence I-50125, Italy.\\
$^4$Harvard-Smithsonian Center for Astrophysics, 60 Garden Street, Cambridge
MA 02138.\\
$^5$Instituto de Astronomia, Universidad Nacional Autonoma de Mexico,\\
Apartado Postal 70-264, Ciudad Universitaria, Mexico, D.F., CP 04510,
Mexico.
}

\date{February 3, 2007}

\pagerange{\pageref{firstpage}--\pageref{lastpage}} \pubyear{2005}

\def\LaTeX{L\kern-.36em\raise.3ex\hbox{a}\kern-.15em
    T\kern-.1667em\lower.7ex\hbox{E}\kern-.125emX}

\begin{document}

\label{firstpage}

\maketitle
\begin{abstract}
We have analyzed the two longest (elapsed time $\gs3$ days)
BeppoSAX observations of the X-ray brightest Seyfert galaxy, NGC 4151,
to search for spectral variability on timescales from a few tens of
ksec to years. We found in both cases highly significant spectral
variability below $\approx 6$ keV down to the shortest timescales
investigated.  These variations can be naturally explained in terms of
variations in the low energy cut-off due to obscuring matter along the
line of sight.  If the cut-off is modeled by two neutral absorption
components, one fully covering the source and the second covering only
a fraction of the source, the shortest timescale of variability of a
few days constrains the location of the obscuring matter to within
3.4$\times 10^4$ Schwarzschild radii from the central X-ray
source. This is consistent with the distance of the Broad Emission
Line Region, as inferred from reverberation mapping, and difficult to
reconcile with the parsec scale dusty molecular torus of
\cite{krbe}. We have also explored a more complex absorption
structure, namely the presence of an ionized absorber. Although the
behaviour of the ionization parameter is nicely consistent with the
expectations, the results are not completely satisfactory from the
statistical point of view.

The overall absorption during the 2001 December observation is lower
than in all other historical observations with similar 2-10~keV
flux. This suggests that absorption variability plays a crucial role
in the observed flux variability of this source.  
\end{abstract}


\section{Introduction}

In the Unified Scheme for AGN (\citealt{ant}, \citealt{urpa}), type~2 narrow
lines Active Galactic Nuclei (AGNs) are normal type~1 AGNs with both the
characteristic broad emission lines and the optical to X-ray continuum
observed through a column of gas and dust of density N$_H\gs10^{22}$
cm$^{-2}$. In this model, the absorption is located in a dusty torus at parsec
distances from the central continuum (\citealt{krbe}, \citealt{pikr},
\citealt{pikr2}). Large variations of the obscuring screen are therefore not
expected on timescales much shorter than the crossing time at the dust
sublimation radius (r$_d$$=$4$\times$L$_{UV,46}^{1/2}$T$^{-2.8}_{1500} \times
10^{18}$cm, where L$_{UV,46}$ is the ultraviolet luminosity in units of
10$^{46}$ erg s$^{-1}$ and T$_{1500}$ is the grain evaporation temperature in
units of 1500 K, \citealt{bar}). However, \citet{riel} found that 23/24 X-ray
absorbed (Compton ``thin'', i.e.N$_H\ls10^{24}$ cm$^{-2}$) AGNs showed N$_H$
variability by a factor 2-3, and, most interestingly, that several objects
varied on timescales of months (but see the case of NGC 4388 for even shorter
variability timescales: \citealt{elri}), implying a limit to the distance r
of the absorber from the X-ray source of $ 4.5 \times 10^{16}
{{M_\bullet}\over{10^9~M_\odot}} ({{n_e}\over{10^6~cm^{-3}}})^{2}
({{N_H}\over{5 \times 10^{22}~cm^{-2}}})^{-2} $ cm. Therefore \citet{riel}
suggested that the absorber might be located in cool clouds present in an
accretion disk wind on the scale of the Broad Emission Line Region
(BELR). This echoes the model of \citet{kako}, which predicts N$_H$
variability down to a timescales of days. Variations of the overall level of
the absorption may also be expected in the case of an ionized absorber
responding to variations of the continuum level (see e.g Nicastro et al. 1999, Schurch
\& Warwick~2002). An accurate analysis of the temporal behaviour of the X-ray
absorber can help us to gain precious information on its physics, size,
geometry and location. Unfortunately, this has been possible so far in a
handful of cases only (e.g. NGC~4388, Elvis et al.~2004, NGC~1365, Risaliti et
al.~2005, Mkn~348, Smith et al. 2001), because of the strong parameter
degeneracy in complex, multi-component spectral fittings, in particular
between the absorber and the continuum parameters. High statistics, broad-band
observations are crucial to remove this degeneracy. To this purpose, we have
analyzed two BeppoSAX long (elapsed time $\gs3$ days) observations of the
bright Seyfert galaxy, NGC 4151, which are particularly well suited to
further investigate this issue. In this paper we focus on the search for
variability of the absorber on timescales from a few tens of ksec to years.

NGC 4151 is one of the brightest Seyfert galaxy in the sky in the X-ray
band, and for this reason it has been studied in detail by all X-ray missions
since the discovery of its X-ray emission in the 1971 \citep{guke}.  The
intrinsic continuum has historically been parameterizered by an absorbed power
law, with variable intensity and photon index ($1.3< \Gamma <1.9 $,
\citealt{ivsa}, \citealt{bawh}, \citealt{pamu}, \citealt{pepi},
\citealt{yawa}, \citealt{fipe}, \citealt{scwa}, \citealt{zdle}). A strong
narrow iron K$_\alpha$ line is present in the spectrum.  This line has an
intensity of 1.8$\times$ 10$^{-4}$ ph cm$^{-2}$ s$^{-1}$ (HETGS
(i.e. High-Energy Transmission Grating Spectrometer, \citealt{cani})
\citealt{ogma}). The narrow core of the line is unresolved, with FWHM$=<$2000
km s$^{-1}$. The Fe K$_\alpha$ intensity remains constant over short-medium
(days-months) timescales and varies by $\sim 25 \%$ on timescales of about
one year (\citealt{pepi}, \citealt{zdle}, \citealt{scwa}, \citealt{tain},
\citealt{scwa2}).  These timescales suggest that the Fe K$_\alpha$ line
arises in a region at a distance r$\sim 10^{17}$ cm from the nucleus. A
Compton reflection component has been detected in the spectrum of NGC 4151 by
 \citet{zdle}, \citet{scwa}. At energies above 50 keV the
continuum becomes steeper, and this change in slope can be modelled by an
exponential cut-off with e-folding energy of 70$-$250 keV (Zdziarski et
al.~1996).

The nuclear power-law spectrum is heavily absorbed at E$\le$ 6 keV by a
photoelectric absorption due to a large amount of material along the line of
sight (N$_H$$\sim$ 10$^{22}$$-$10$^{23}$ cm$^{-2}$). A single absorber does
not provide a good fit, as recognized since the first Einstein observations
(SSS, \citealt{homu}). A better description of the spectrum is provided by an
inhomogeneous absorber, which allows some fraction of the nuclear X-ray
continuum to emerge with significantly less attenuation than the rest of the
continuum. Both a partial covering absorber (\citealt{homu}, \citealt{pepi},
\citealt{zdle}) and an ionized absorber (\citealt{yawa}, \citealt{weya},
\citealt{scwa}) have been proposed in the past. Part of the emission is due to
extended kpc-scale emission associated with the narrow line region
(\citealt{elbr}, \citealt{ogma}).  Interestingly, changes of the absorbing
screen(s) on timescales of days to years have been often reported in the past
(\citealt{bawh}, \citealt{fipe}, \citealt{yawa2}, Schurch \& Warwick~2002).
These changes have been interpreted in terms of creation and destruction of
cold filaments and clouds in the BELR \citep{bawh}, or in terms of their
orbital motion (\citealt{homu}, \citealt{lael}).  In the latter case, the
inhomogeneous absorber consists of a large number of discrete clouds, each
much smaller than the size of the X-ray source. The number of the clouds that,
at any time, cover a given line of sight is distributed according to the
Poisson statistics. Column density variations, in this scenario, are simply
due to statistical fluctuations of the number of clouds along the line of
sight. The thickness and size of the absorbing clouds can therefore be
estimated by measuring amplitude and timescales of the column density
variations. In models of a partially ionized absorber the variations in the
underlying ionizing continuum change the optical depth of the absorber along
the light of sight and so simulate a change in equivalent H column density,
when the spectra are modelled with a neutral absorber. In this scenario,
Schurch \& Warwick (2002) interpret the spectral variability observed in a
ASCA long look in terms of changes of the ionization state in response of
variations of the underlying continuum flux.  This interpretation relies on
the assumption of a constant underlying continuum shape, evaluated using a
broad-band BeppoSAX observation. In this paper we report on further BeppoSAX
observations which reveal changes in the obscuring column density on
timescales of a few days, which strongly constrain the geometry of the
absorber.  Thanks to the high statistics and broad spectral coverage we were
able for the first time to investigate the nature of the observed variability
without resorting on any a-priori assumption on the continuum emission.

The structure of this paper is the following: Section 2 gives some
details on the data reduction, Section 3 illustrates the method used
for the data analysis and presents the results of the temporal
and spectral analysis, Section 4 is devoted to the discussion of our
results and  conclusions.

\section{Observations and data reduction}

To look for changes in the absorber(s) of NGC~4151 and to further
investigate on its origin and geometry, we have selected the two
longest observations performed with the BeppoSAX Narrow Field
Instruments (Table 1), LECS (0.1-10 keV, \citealt{pama}), MECS (1.3-10
keV, \citealt{boch}) and PDS (13-200 keV, \citealt{frco}), which have
good sensitivity from 0.1 to at least 100 keV. Because of the
complexity of the spectrum, an energy coverage as broad as possible is
the key to obtain a good constrain on the absorption models.
The first observation was performed between 1996 July 6 and 9 for a
total elapsed time of $\sim300$ ksec.  During the observation the
2-4 keV flux increased by a factor $\sim$3, and therefore this
observation is particularly well suited to search for spectral
variability. The second observation was a target of opportunity
observation (ToO) performed between 2001 December 18 and 20, for a
total elapsed time of $\sim275$ ksec, during a NGC~4151
outburst in the soft X-rays. This outburst was serendipitously
detected by the BeppoSAX Wide Field Cameras, and a ToO observation of
the source with the Narrow Field Instruments was promptly
performed. In this observation the signal to noise is particularly
good and allows us to search for small spectral variation, and for variation
on short timescales.

The 1996 July observation was performed with MECS units 1, 2 and 3,
while the 2001 December observation was performed with MECS units 2
and 3 only (on 1997 May 6 a technical failure caused MECS unit 1 to be
switched off); data from the units were combined after gain
equalization. The LECS was operated during dark time only, therefore
LECS exposure times are smaller than MECS ones.  Table 1 gives the
LECS, MECS and PDS exposure times and the mean count rates.

\begin{table*}
\begin{minipage}{126mm}
\caption{ Observation log.}
\begin{tabular}{lcccc}
\hline
Instrument &channels  & Energy range&  Net counts s$^{-1}$ &  Exposure (s)\\
\hline
1996 July 6-9\\
\hline
LECS &  41-370    & 0.45-4 keV    & 0.109$\pm$0.003  &    8960    \\
MECS & 37-227     & 1.65-10.5 keV  &  1.446$\pm$0.004   &   73620    \\
PDS  &  50-580   & 15-200 keV   &  4.09$\pm$ 0.03  &   35160   \\
\hline
2001 December 18-20\\
\hline
LECS &  41-370      & 0.45-4 keV   &    0.395$\pm$0.003   &   43300     \\
MECS &   37-227    &   1.65-10.5 keV &   2.070$\pm$0.004  &  114860       \\
PDS  &  50-580      & 15-200 keV   &  5.10$\pm$0.02   &   53290   \\
\hline
\end{tabular}

\medskip
\end{minipage}
\end{table*}

Standard data reduction was performed using the SAXDAS software
package version 2.0 following \citet{figu}.  In particular, data are
linearized and cleaned of Earth occultation periods (we accumulated
data for Earth elevation angles $>5$ degrees) and unwanted periods of
high particle background (satellite passages through the South
Atlantic Anomaly and periods with magnetic cut-off rigidity $>6$
GeV/c).

NGC 4151 lies 5 arcmin south of the relatively bright BL Lac object
1E1207.9+3945 (0.3-3.5 keV flux of $1.5\times 10^{-12}$ \cgs,
\citealt{magi}) and $\sim$5.2 arcmin south of the galaxy NGC 4156
(which is $\sim$20 times fainter than NGC 4151, \citealt{elvis2},
\citealt{faki}). To study the possible contamination from these two
sources we compared the LECS and MECS spectra extracted from regions
of 2, 3 and 4 arcmin radii.  The spectra are all consistent with each
other in shape and therefore we conclude that the contamination of the
nearby sources is negligible, even in the 4 arcmin radius
spectra. Since these spectra provide the best signal to noise, we
use them in the following analysis.

Background spectra were extracted in detector coordinates from
high Galactic latitude `blank' fields (98\_11 release) using
regions equal in size to the source extraction region. We have
compared the mean level of the background in the LECS and MECS
``blank fields'' observations to the mean level of the background
in the NGC 4151 observations using source free regions at various
positions in the detectors. The ``local'' MECS, LECS background
count rates are within a few per cent of those in the ``blank
fields''.

The PDS data were reduced using the ``variable risetime threshold''
technique to reject particle background (see \citealt{figu}). To check
the reliability of the background subtraction in the PDS spectra we
looked at the spectrum between 200 and 300 keV, where the effective
area of the PDS to X-ray photons is small and therefore the source
contribution is negligible. After background subtraction we obtain a
count rate of 0.01$\pm$0.01 counts s$^{-1}$, for the 1996 July
observation and 0.003$\pm$0.009 counts s$^{-1}$ for the 2001 December
observation, fully consistent with the expected value of 0. The BL Lac
object 1E1207.9+3945 and the galaxy NGC 4156 contribute less than a
few per cent to the PDS flux, based on the extrapolation of the 2-10
keV spectrum of these sources to higher energies, assuming a power law
spectrum with $\Gamma=2.1$ \citep{pest} and a 7 keV exponential
spectrum \citep{bujo} respectively. NGC 4151 is detected with a signal
to noise ratio $S/N\gs3$ up to 130$-$140 keV in the two observations.

Spectral fits were performed using the XSPEC 11.2.0 software
package and the most recent responses (1999 December release). For
both LECS and MECS we used the standard on-axis responses, since
the source is close to the default pointing position (within 5 raw
pixels, i.e. 40 arcsec).  LECS and MECS spectra were binned
following two criteria: (a) allowing at least four channels per
resolution element at all energies, when possible, and (b) to
obtain at least 20 counts per energy channels.

Constant normalization factors have been introduced in the fitting
models in order to take into account the intercalibration
systematics between the instruments \citep{figu}. The
normalization factor between the LECS and the MECS instruments,
assuming the MECS as a reference, has been obtained by fitting the
LECS and MECS spectra in the common 1.65-4 keV band with a simple
power law model. We obtained a normalization factor of
0.40$\pm0.02$ for the 1996 July observation and a factor of
0.60$\pm0.01$ for the 2001 December observation. Accordingly, in
all the following fits the LECS-MECS factor is constrained to vary
in the above ranges for the 1996 July observation and for the 2001 December
observation respectively. The LECS-MECS
normalization factor is somewhat lower than usually assumed (see
\citealt{figu}), due to the adopted LECS source extraction region
of radius 4 arcmin, smaller than the typical one (because of
confusion problems, see above). The normalization factor adopted
between the PDS and the MECS instruments is constrained to vary between 0.72
and 0.87, as recommended  (see \citealt{figu}).

In the spectral fits we used the LECS between 0.45-4 keV (channels
41-370), MECS between 1.65-10.5 keV (channels 37-227), and the PDS
between 15-200 keV (channels 50-580). Errors throughout the paper are quoted at a
significance level of 90\% for two interesting parameters ($\Delta
\chi^2 = 4.61$ \citealt{lama}), unless differently specified.

\section{Data analysis}

The aim of this paper is to search for variations of the
absorber(s) on timescales from a few tens of ksec to years.  The
total absorbing column density measured toward NGC~4151 ranges from
a few $10^{22}$ cm$^{-2}$ to a few $10^{23}$ cm$^{-2}$ and the
corresponding photoelectric cut-off lies in the 2-4 keV energy
range. This band is therefore best suited to monitor changes in
the absorber. The 6-10 keV and 15-100 keV bands, being less
affected by photoelectric absorption can be used to monitor the
continuum, normalization and power law photon index.

\subsection{Light curves analysis}

\begin{figure}
\centering
\includegraphics[angle=0,height=16truecm,width=10truecm]{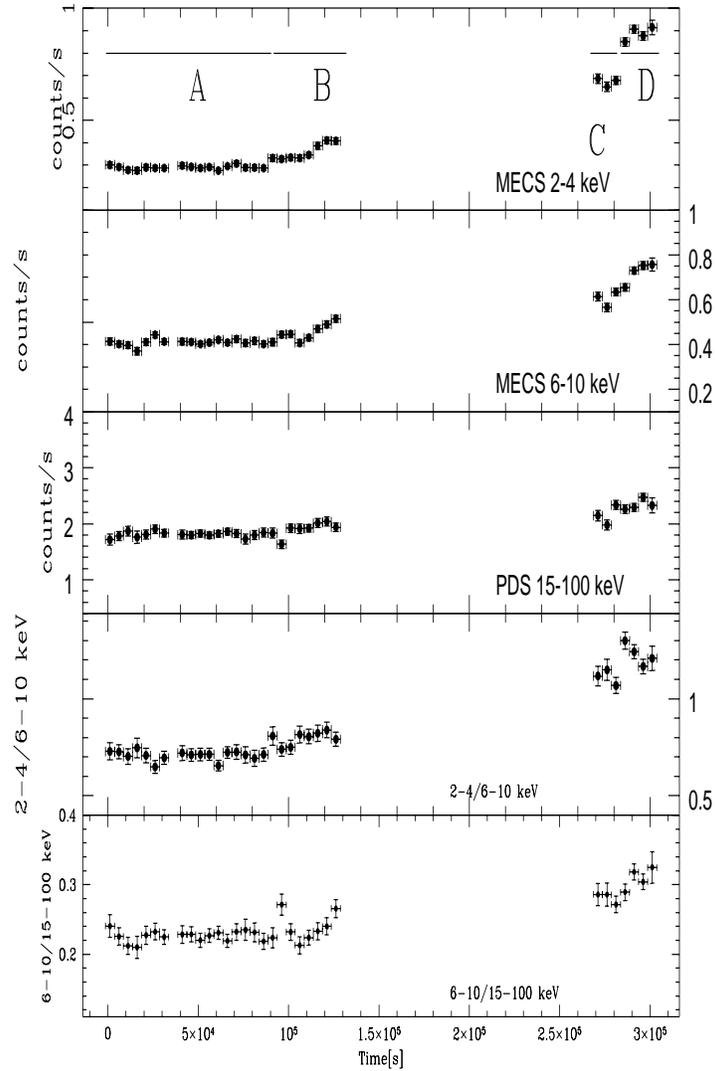}
\caption{Light curves of the 1996 July observation in bins of 5500
seconds ($\sim$ 1 satellite orbit). From top to bottom: MECS 2-4 keV
count rate, MECS 6-10 keV
count rate, PDS 15-100 keV half detector count rate, 2-4 keV/6-10 keV softness
ratio, 6-10 keV/15-100 keV softness ratio. Capital letters and lines
in the top panel indicate the time intervals selected for a
time-resolved spectral analysis.}
\label{lc_july96}
\end{figure}

\begin{figure}
\centering
\includegraphics[angle=0,height=16truecm,width=10cm]{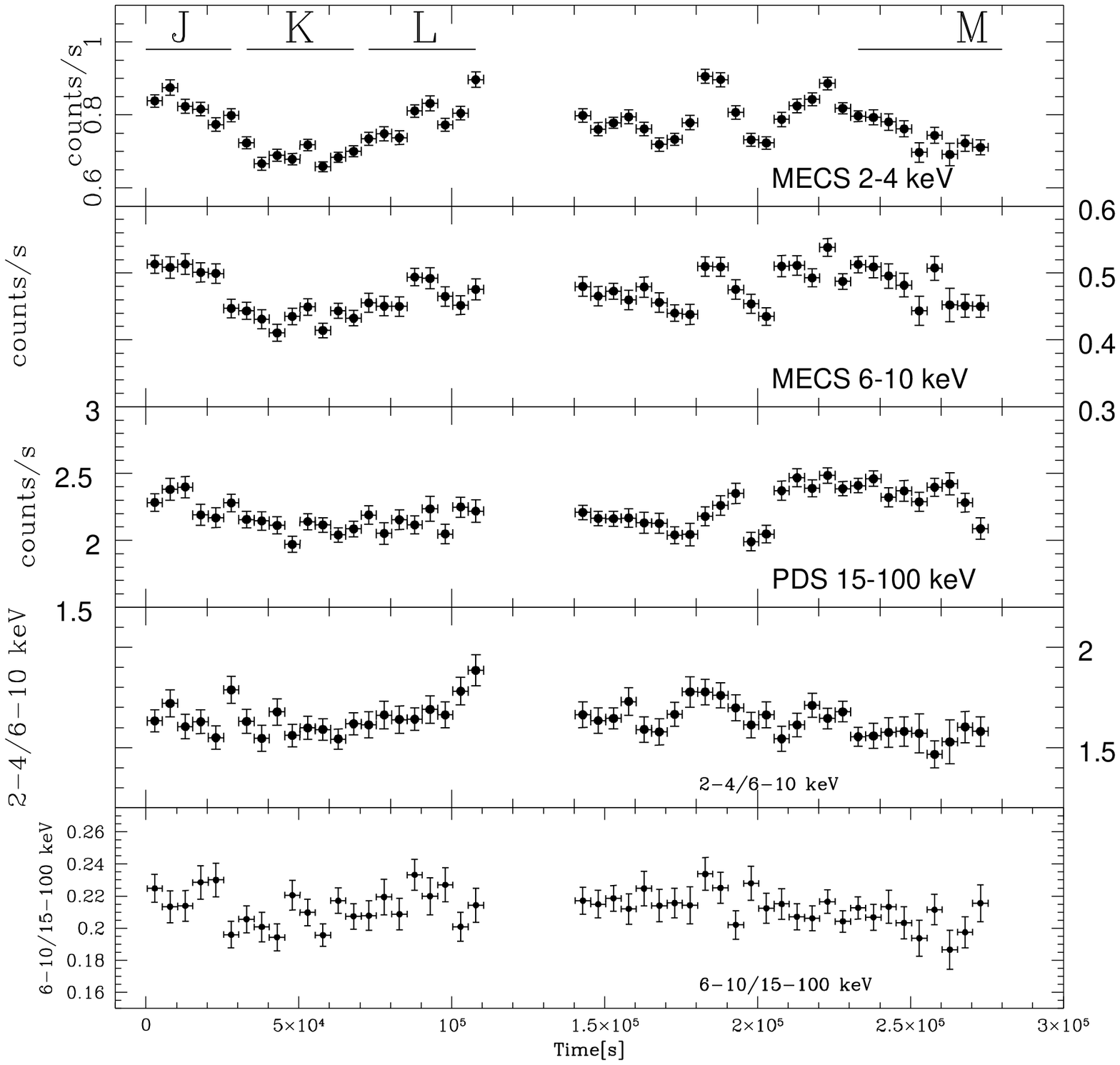}
\caption{Light curves of 2001 December observation in bins of 5500
seconds ($\sim$ 1 satellite orbit).  From top to bottom: MECS 2-4 keV
count rate, MECS 6-10 keV
count rate, PDS 15-100 keV half detector count rate, 2-4 keV/6-10 keV softness
ratio, 6-10 keV/15-100 keV softness ratio. Capital letters and lines
in the top panel indicate the time intervals selected for a
time-resolved spectral analysis.}
\label{lc_dec01}
\end{figure}

Light curves of the count rate in the three energy bands 2-4 keV, 6-10
keV, 15-100 keV, and the softness ratios 2-4 keV/6-10 keV and 6-10
keV/15-100 keV, are plotted in figures \ref{lc_july96} and
\ref{lc_dec01} in bins of 5500 seconds ($\sim$ 1 satellite orbit).
The analysis of both count rate and softness ratio light curves
indicates large spectral variations on timescales from a few tens of
ksec to a few days. 

\begin{figure}
\centering
\includegraphics[angle=0,height=8truecm,width=10truecm]{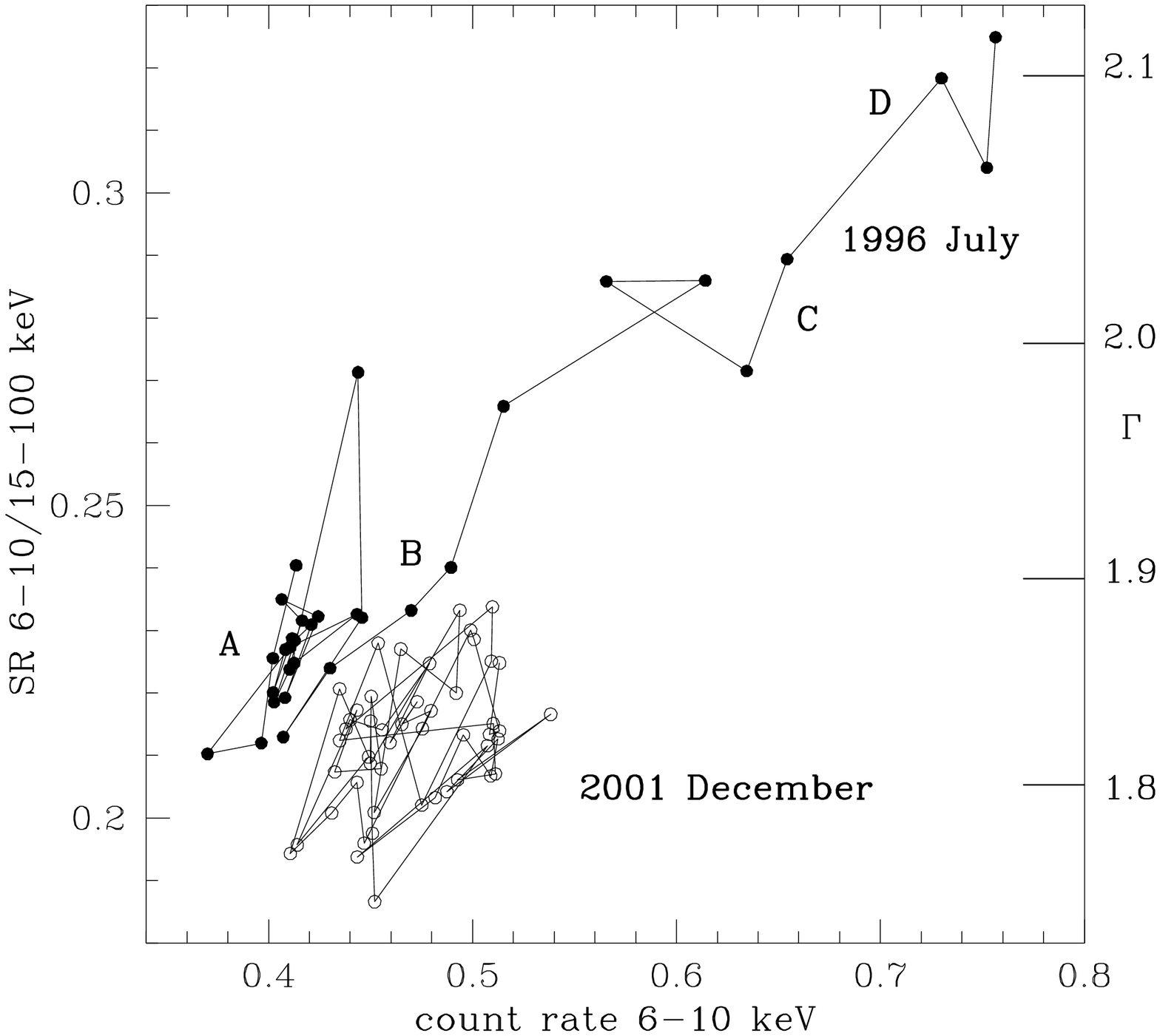}
\caption{6-10 keV/15-100 keV softness ratio versus the count rate in
6-10 keV range . The open dots indicate the 2001 December observation,
the solid dots indicate the 1996 July observation. The lines on the
right hand side are the 6-10 keV/15-100 keV softness ratio expected
from a power law model with photon index $\Gamma$, reduced at low energy by
photoelectric absorption from neutral gas of column density
N$_H$=10$^{23}$ cm$^{-2}$.}
\label{hr_cr}
\end{figure}

\begin{figure}
\centering
\includegraphics[angle=0,height=8truecm,width=10truecm]{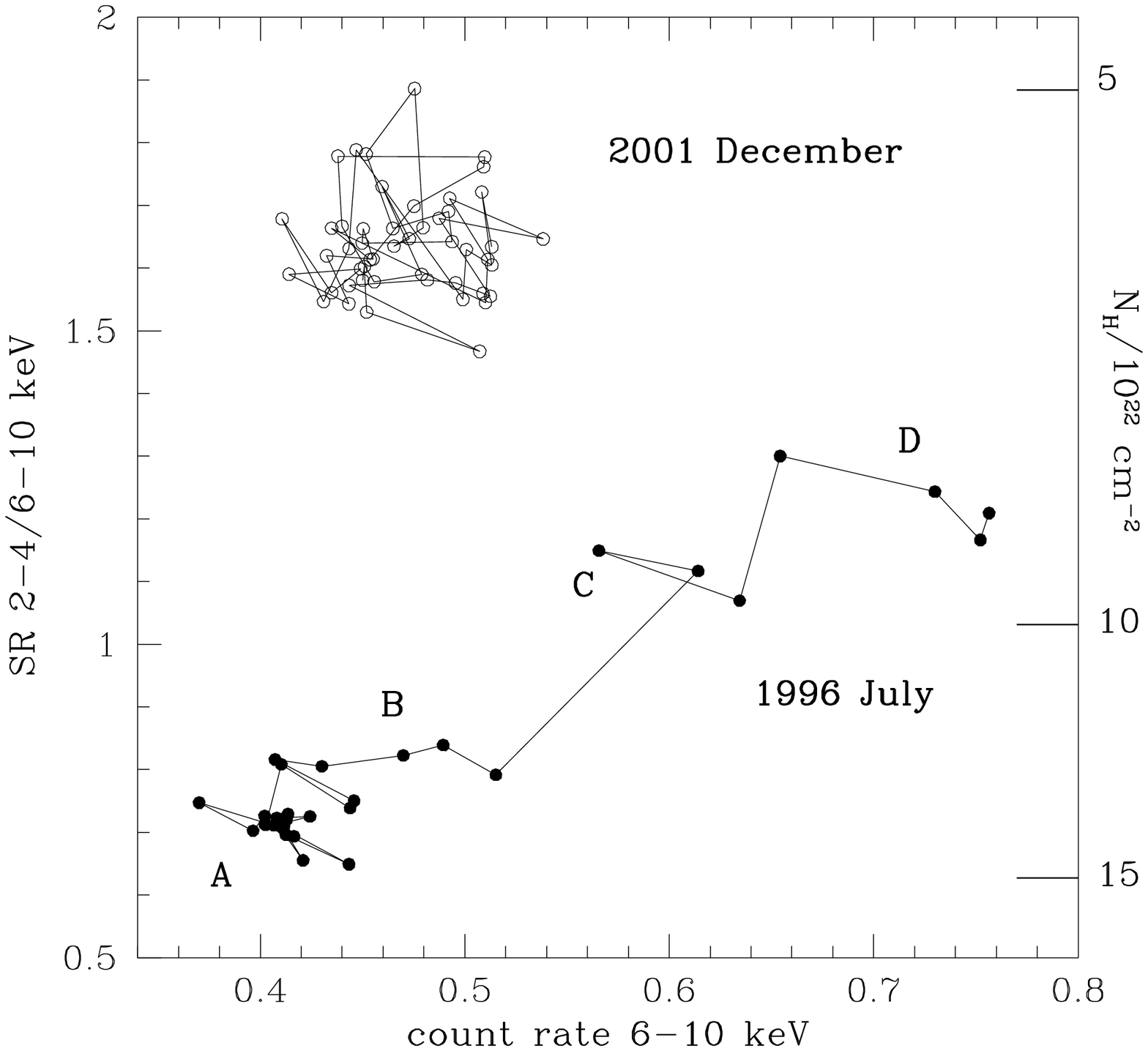}
\caption{2-4 keV/6-10 keV softness ratio versus the count rate in
6-10 keV range . The open dots indicate the 2001 December
observation, the solid dots indicate the 1996 July observation.The lines on the
right hand side are the 2-4 keV/ 6-10 keV softness ratio expected
from a power law model with photon index 1.8, reduced at low energy by
photoelectric absorption from neutral gas of column density N$_H$.}
\label{sr_cr}
\end{figure}

On longer timescales (i.e. the 4.5 years between the two observations), we
find similar results.The 6-10 keV count rates in the intervals A and B of the
1996 July observation are similar to the corresponding count rates recorded
during the 2001 December observation (see figures \ref{hr_cr}), as well as the
the 6-10 keV/15-100 keV softness ratios.  Conversely, the 2-4 keV/6-10 keV
softness ratios are different by a factor $\sim$2 (see figure
\ref{sr_cr}). The similarity of the higher energy count rates (and 6-10 keV/
15-100 keV softness ratios) suggests a comparable photon index in the range
$\sim$1.75-1.85 (assuming a simple power law model reduced at low
energy by a uniform column of cold gas). On the other hand, the difference in
the 2-4 keV/6-10 keV ratios can be naturally explained by secular changes of
the absorber of $\gs 60 \%$.

\subsection{Spectral analysis}

\begin{figure}
\centering
\includegraphics[angle=0,height=8truecm,width=10truecm]{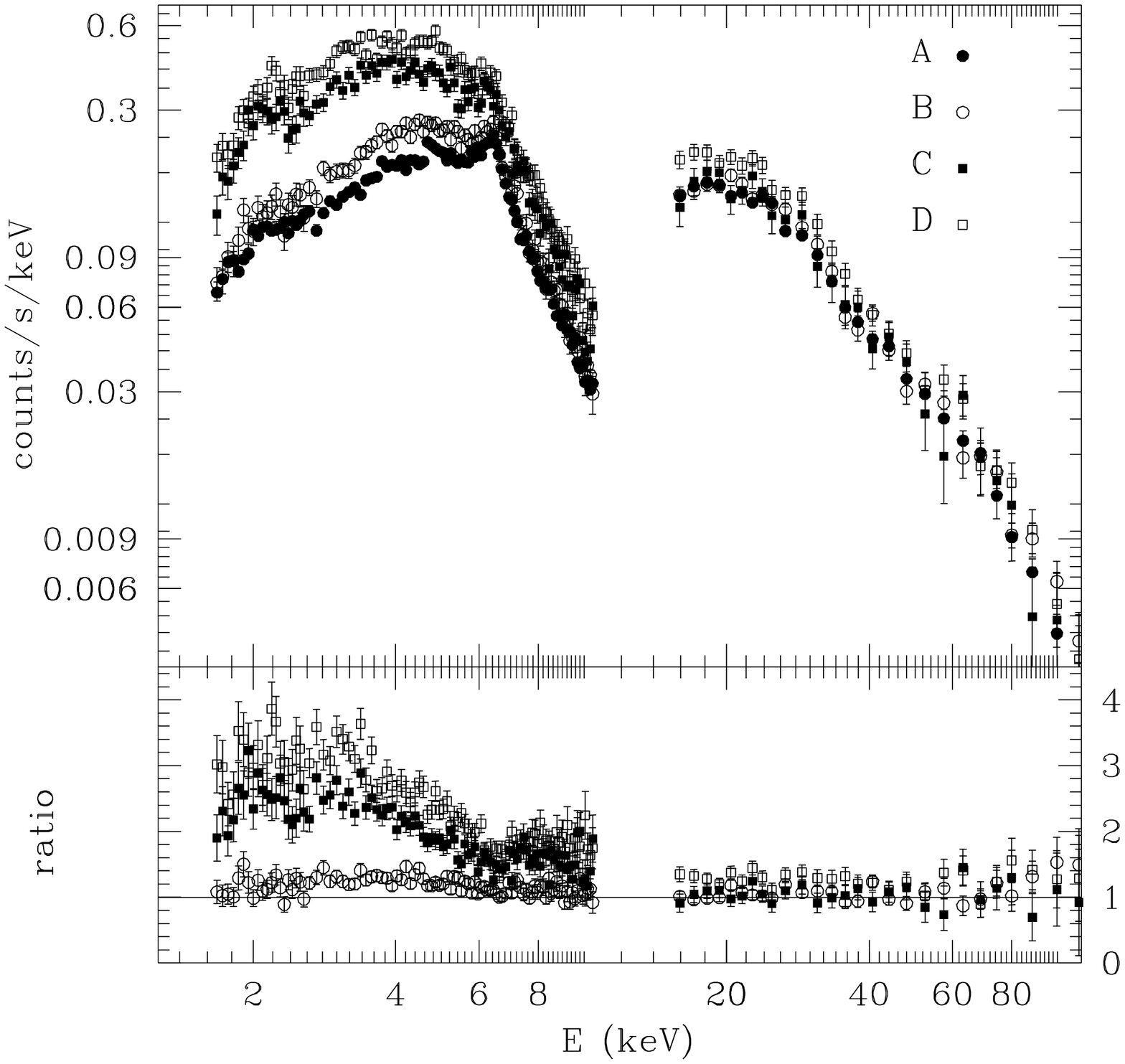}
\caption{1996 July observation. Top panel: counts spectra of the
data in the selected time intervals indicated in figure
\ref{lc_july96}. Bottom panel: ratio between the counts spectra to
the lower counts spectrum, A. LECS data are not shown for
the sake of clarity (statistical scatter is comparable with the observed
variability).} \label{spe_6}
\end{figure}

\begin{figure}
\centering
\includegraphics[angle=0,height=8truecm,width=10truecm]{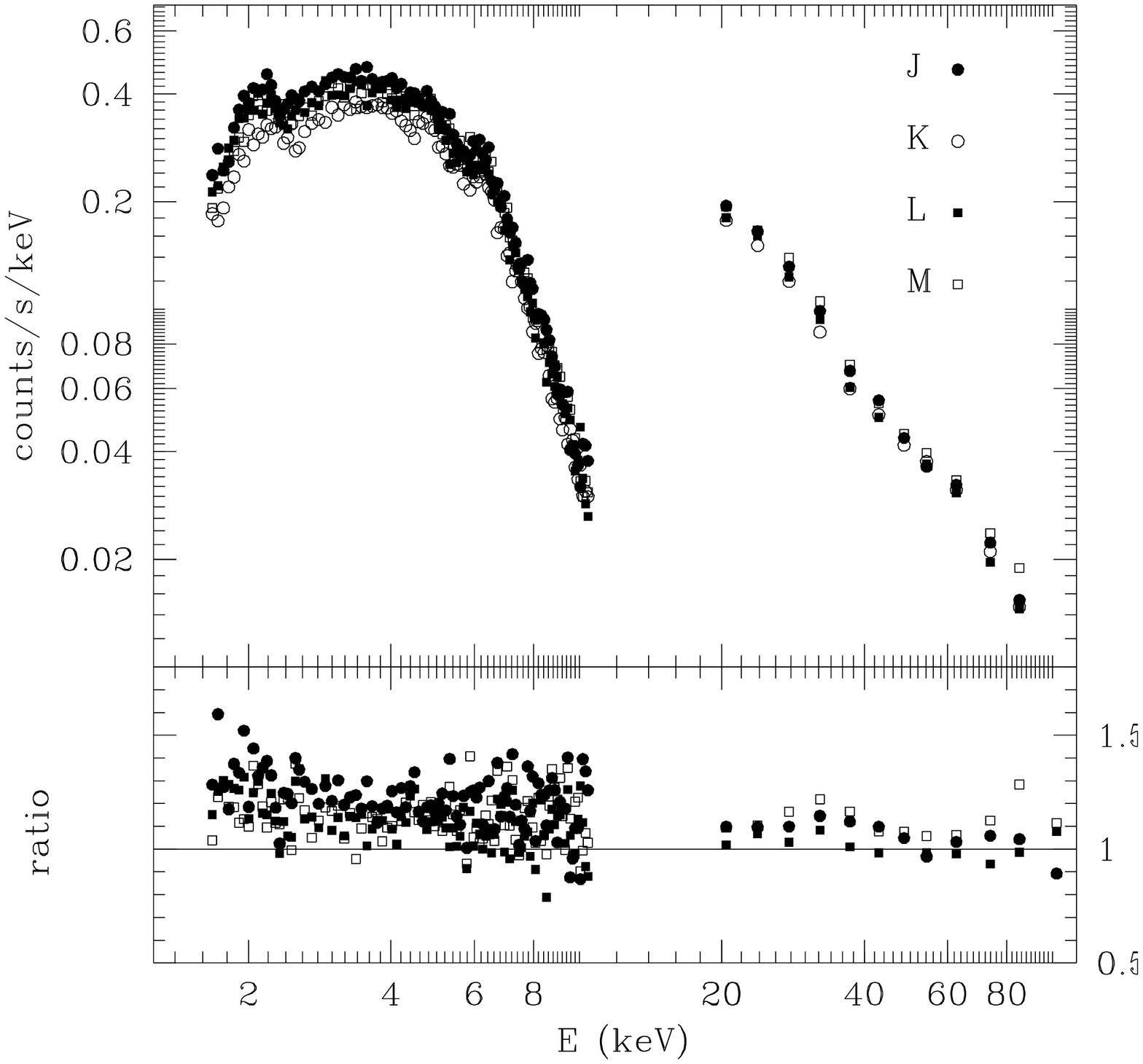}
\caption{2001 December observation. Top panel: counts spectra of
the data in the selected time intervals indicated in figure
\ref{lc_dec01}.  Bottom panel: ratio between the counts spectra to
the lower counts spectrum, K. LECS data are not shown for
the sake of clarity (statistical scatter is comparable with the observed
variability).} \label{spe_18}
\end{figure}

The statistics of the spectra extracted in 5500 s bins is not high
enough to constrain complex models. To characterize the spectral
variability we therefore accumulated spectra in contiguous time
intervals when the source does not display large spectral variations.
We selected four time intervals for each of the two observations to
perform a time resolved spectral analysis (indicated as A, B, C and D
in figure \ref{lc_july96} and as J, K, L, M in figure \ref{lc_dec01}).
The counts spectra in these time intervals are plotted in the upper
panels of figures \ref{spe_6} and \ref{spe_18}.  The lower panels plot
the same spectra normalized to the lower count rate spectrum of each
observation. The largest variations are seen below $\approx 6$ keV, 
thus suggesting variations of the absorber(s).

The comparison of figures \ref{spe_6} and \ref{spe_18} suggests that
the absorber variations during the 2001 December observation are
smaller than those in the 1996 July observation, but qualitatively
similar.
To fully characterize the observed spectral variations we performed a
detailed spectral analysis of the spectra plotted in figures
\ref{spe_6} and \ref{spe_18}.  We have adopted a model including the
following components. The hard X-ray continuum is described by a
power-law with an exponential high-energy cut-off plus a neutral
Compton reflection component (the {\sc PEXRAV} model in {\sc XSPEC},
\citealt{mazd}). The latter component can be safely assumed constant on timescales
from days to months (see e.g. Johnson et al. 1997) and therefore its
normalization has been fixed to the best fit value found fitting the
spectra of each complete observation.  A narrow iron K$_\alpha$
emission line with energy set to 6.4 keV (rest frame) is added to this
continuum.

The soft X-ray continuum is described by two components, a thermal
bremsstrahlung component (\citealt{elfa}, \citealt{weya} and references
therein), and a scattering component \citep{weya}. The thermal component can
be safely assumed constant on the timescales spanned by our observations and
therefore we fixed temperature and normalization to the best fit values found
by fitting the total 2001 December spectrum (T$=$0.18 keV and a 0.4-4 keV flux
of $1.8\times 10^{-12}$ erg cm$^{-2}$ s$^{-1}$).  The low energy scattering
component, parameterizered by a power law, is likely produced by scattering of
the nuclear power law by warm-hot gas distributed on scales greater than
parsec (\citealt{ogma}, \citealt{yawi}).Therefore, also this component can be
safely assumed constant on the timescales spanned by our observations. We
fixed its photon index to 1.7 and the normalization to the best fit values
found by fitting the total spectra (0.4-4 keV flux of $6.1\times 10^{-12}$ erg
cm$^{-2}$ s$^{-1}$ and of 4.5$\times 10^{-12}$ erg cm$^{-2}$ s$^{-1}$ for the
1996 July and the 2001 December observations, respectively).

The complex absorber has been parameterizered in three different ways:
A) two neutral components, one covering the nucleus totally and the
other covering the nucleus only partly; B) one neutral absorber and
one ionized absorber both covering the nucleus totally; and C) one
ionized absorber covering the nucleus totally and one neutral absorber
covering the nucleus only partly.  A Galactic column density along the
line of sight of N$_H (gal)=2.1 \times 10^{20}$ cm$^{-2}$ (Murphy et
al.  1996) has been added to the model in all cases.

Models A) and B) have seven free parameters: the photon index $\Gamma$,
normalization and the energy of the cut-off of the nuclear continuum,
three parameters for the absorber (N$_H$, N$_H$c, c$_v$, or N$_{H1}$,
N$_{Hw}$, and the ionization parameter $\xi$ ) and the normalization
of the iron K$_\alpha$ line.  Model C) has one more free parameter
(the covering fraction of the neutral absorber).

\begin{figure}
\centering
\includegraphics[angle=0,width=15cm]{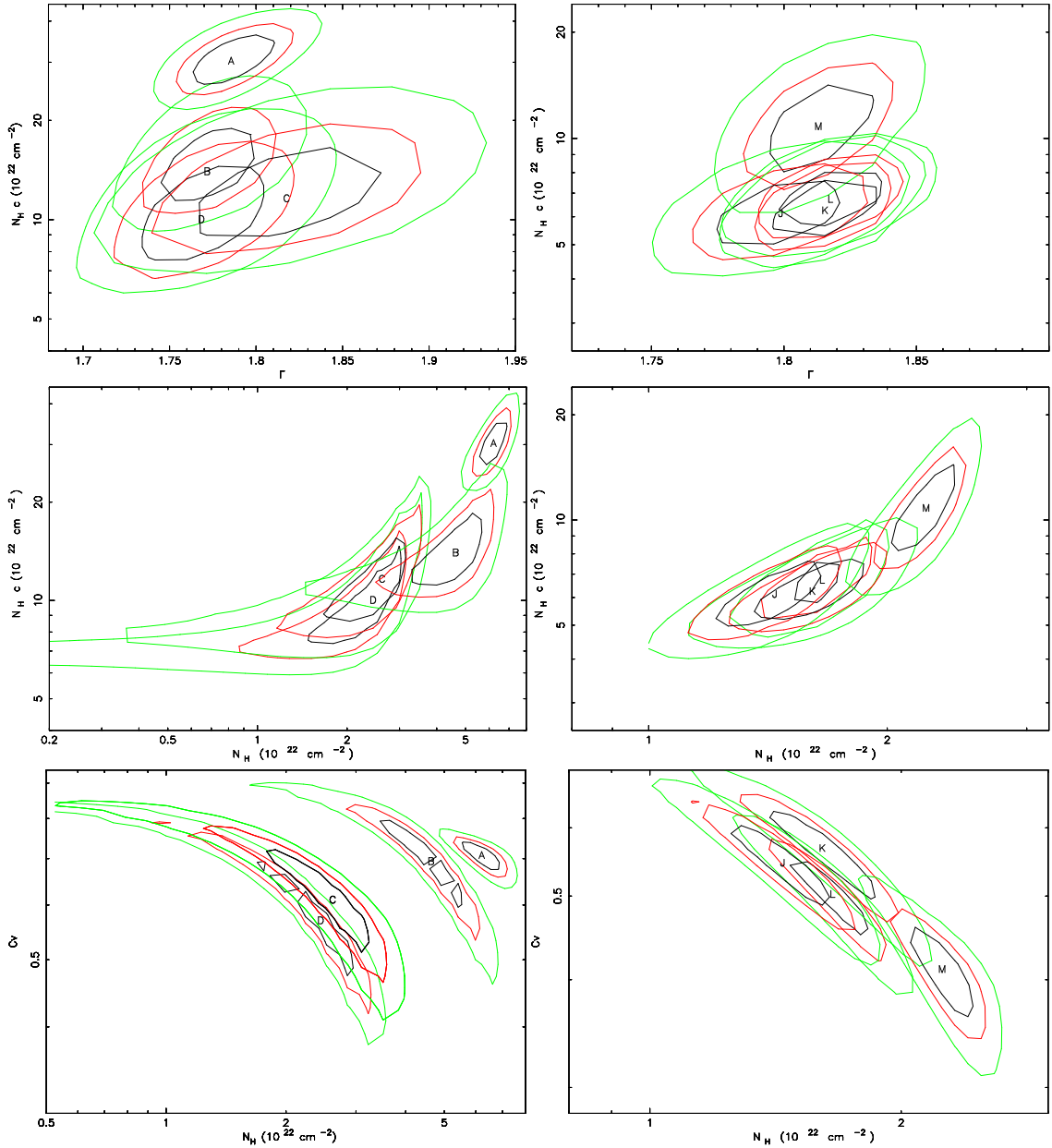}
\caption{68\%, 90\%  and 99\% $\chi^2$ confidence
contours  of the partial covering absorber versus the photon index (upper panels),
partial
covering absorber versus the totally covering absorber (central panels) and
partial
covering factor versus the totally covering absorber (lower panels)
for the 1996 July observation (left panels) and the 2001 December observation
(right panels).}
\label{gn2_6}
\end{figure}

Model A) provided the smallest $\chi^2$ in both 1996 July and 2001 December
observations. We then discuss first the results of this series of fits and
then compare them with the results obtained using models B) and C).  Tables 2
and 3 give the best fit values of N$_H$, N$_H$c, c$_v$, $\Gamma$, along with
the $\chi^2$, the count rate and the flux in the 6-10 keV band, for the four
spectra selected in each observation.  In all cases the fits are acceptable at
a confidence level better than $3\%$.  Figure \ref{gn2_6} shows the 68\%,
90\%, and 99\% $\chi^2$ confidence contours of N$_H$c versus $\Gamma$, N$_H$
versus $N_H$c and C$_v$ versus N$_H$ for the 1996 July and the 2001 December
observations.

\begin{table*}
\begin{minipage}{126mm}
\caption{1996 July observation: the best fit parameters of model A.}
\begin{tabular}{lccccccc}
\hline
spectrum & $\chi^2$  & N$_H$$^a$ & N$_H$c$^b$ & c$_v$$^c$ &
$\Gamma$$^d$  & c$_{6-10}$$^e$& F$_{6-10}$$^f$ \\
\hline
A & 149.20/151 &   6.2$\pm_{1.0}^{1.0}$  &  30.3$\pm_{6.5}^{9.0}$  &
0.71$\pm_{0.05}^{0.04}$  &  1.79$\pm_{0.04}^{0.03}$ &  0.41& 5.5\\

B & 147.74/138  &  4.6$\pm_{2.1}^{1.7}$ & 14.0$\pm_{3.7}^{8.0}$ &
0.7$\pm_{0.2}^{0.2}$  & 1.77$\pm_{0.04}^{0.04}$ & 0.45& 6.0 \\

C &  119.10/118 & 2.6$\pm_{1.5}^{1.0}$ & 11.6$\pm_{4.0}^{8.1}$ &
0.6$\pm_{ 0.1}^{0.2}$ &  1.82$\pm_{0.08}^{0.04}$ &  0.64 & 8.2\\

D &  124.29/118 &  2.4$\pm_{1.6}^{0.9}$ & 10.0$\pm_{3.5}^{7.3}$ &
0.6$\pm_{0.2}^{0.2}$ & 1.77$\pm_{0.05}^{0.03}$ &  0.73& 9.7 \\

\hline

\end{tabular}

\medskip
$^a$ column density of the totally covering absorber in unit of
$10^{22}$cm$^{-2}$, $^b$ column density of the partial covering absorber in
unit of $10^{22}$cm$^{-2}$,$^c$ covering factor, $^d$ photon index,
$^e$ 6-10 keV count rate, $^f$ MECS 6-10 keV flux in 10$^{-11}$ \cgs.
\end{minipage}
\end{table*}

\begin{table*}
\begin{minipage}{126mm}
\caption{ 2001 December observation: the best fit parameters of model A.}
\begin{tabular}{lccccccc}
\hline
spectrum & $\chi^2$  & N$_H$$^a$ & N$_H$c$^b$ & c$_v$$^c$ & $\Gamma$$^d$
& c$_{6-10}$$^e$ & F$_{6-10}$$^f$ \\
\hline
J &   134.22/136 & 1.4$\pm_{0.3}^{0.3}$ & 6.1$\pm_{1.6}^{2.3}$
& 0.5$\pm_{0.1}^{0.1}$  & 1.80$\pm_{0.04}^{0.04}$ &  0.50 &10.0\\

K & 134.43/136& 1.6$\pm_{0.3}^{0.4}$  & 6.3$\pm_{1.5}^{2.4}$
& 0.6$\pm_{0.1}^{0.1}$  & 1.82$\pm_{0.03}^{0.02}$ &  0.43&8.5 \\

L & 168.93/136& 1.6$\pm_{0.3}^{0.3}$  & 6.7$\pm_{1.6}^{2.3}$
& 0.50$\pm_{0.04}^{0.09}$  &  1.82$\pm_{0.03}^{0.03}$ & 0.46&9.2 \\

M &  152.55/135& 2.2$\pm_{0.3}^{0.3}$  & 10.8$\pm_{3.7}^{5.6}$
& 0.41$\pm_{0.07}^{0.08}$  & 1.81$\pm_{0.03}^{0.03}$  & 0.49&9.7\\

\hline

\end{tabular}

\medskip
$^a$ column density of the totally covering absorber in unit of
$10^{22}$cm$^{-2}$, $^b$ column density of the partial covering
absorber in unit of $10^{22}$cm$^{-2}$, $^c$ covering factor, $^d$
photon index, $^e$ 6-10 keV count rate, $^f$ MECS 6-10 keV flux in 10$^{-11}$ \cgs.
\end{minipage}
\end{table*}

During both observations the photon index is constant at a confidence level
better than 68\% while both absorbers undergo significant changes. During the
1996 July observation N$_H$c and N$_H$ both change by $\sim 60 \%$ at a
confidence level better than $99\%$ from A to C.  Significant (at a confidence
level better than 90 \%) but smaller changes are also present from spectra A
to B.  On the other hand, the covering factor c$_v$ is consistent with a
constant value within the relatively large errors, expecially for spectra C
and D.  During the 2001 December observation we detect significant changes of
the absorbers, despite smaller flux variations.  N$_H$c and N$_H$ change at a
confidence level better than $90\%$ by $\sim 60 \%$ and $\sim 35\%$ from L to
M respectively.  The covering factor changes by 25\% at a confidence level
better than 90\% from J to M.

\begin{figure}
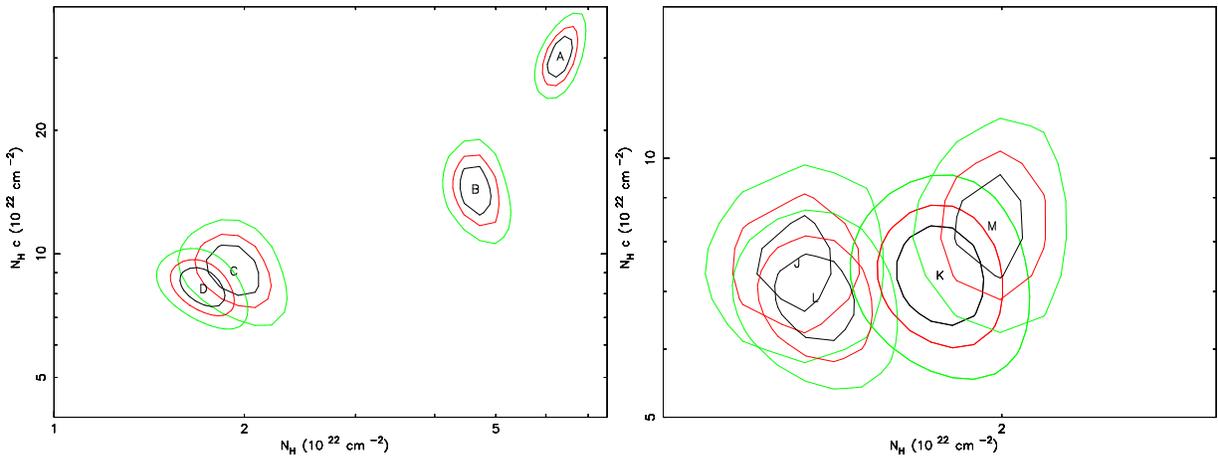

\centering
\begin{tabular}{cc}
\includegraphics[angle=-90,width=8cm]{fig8.ps}
\includegraphics[angle=-90,width=8cm]{fig9.ps}
\end{tabular}
\caption{68\%, 90\% and 99\% $\chi^2$ confidence contours of the partial
covering absorber versus the totally covering absorber, from the fit
with photon index and covering factor frozen. Left panel:1996 July
observation; right panel:2001 December observation.}
\label{n1n2f_6}
\end{figure}

We remark that statistically significant changes of the absorber parameters
have been obtained leaving the power law continuum parameters free to vary and
therefore we can exclude that they are spurious results of subtle variations
of the continuum shape. Having assessed this, we can now fix the
continuum shape so as to obtain better constraints on the absorber
parameters.  Since the covering factor turned out to be constant or limited
within a narrow range, we also fix it in the next series of fits.  These new
constraints do not worsen the fit, the total $\chi^2$ increasing by 14 for 16
additional degrees of freedom.  Figure \ref{n1n2f_6} shows the N$_H$ -- $N_H$c
$\chi^2$ confidence contours for the two observations for this series of fits.
The magnitude of the variations of both parameters is similar to those
reported above, but of course now the $\chi^2$ contours are smaller, and
therefore the variations are more statistically significant. In the July 1996
observation the variations of N$_H$ and N$_H$c seems correlated one to the
other (figure \ref{n1n2f_6}). This may appear unexpected, but it should be
considered that real absorbers would have smooth edges and therefore N$_H$
would vary smoothly, covering the source with a continuous set of N$_H$
values. If this is the case, and if there has been a large variation of
the denser absorber N$_H$c, it is not surprising to find a correlation
between the two parameters of our simple representation of this complex
structure.

In addition to the partial covering model (A), we now explore two versions of
a more complex absorption structure, namely the presence of an ionized
absorber (models B and C).

Tables 4 and 5 report the best fit parameters of model B, obtained
letting the continuum photon index free to vary and fixing it to its
average value (the photon index is again consistent with a constant
value during each observation).

The quality of the fit using model B is worse than that of model
A at a high level of confidence for spectrum A ($\Delta \chi ^ 2 \sim
80$). It is worse for spectrum C and J, K, L, M ($\Delta \chi ^ 2 \sim
4-20$) and it is similar only for spectrum B. The analysis of the
residuals shows that warm absorber models systematically underpredict the
observed spectrum between 1.7 and 2.5 keV. Of course this effect is more prominent
in spectrum A.  We used a simple equilibrium photoionization model {\sc
ABSORI} in {\sc XSPEC}. Conversely, Schurch \& Warwick (2002) argues that the
variable absorber in NGC 4151 is often in a non-equilibrium ionization state.
This may help in explaining the poor fits obtained with {\sc ABSORI}.
Interestingly, the ionization parameter of spectrum D of the 1996 July
observation, when the source was in a high continuum state, is 2-10 times
higher than that of spectra A and B, when the source was in a low state,
although the errors are rather large. This behaviour is fully consistent with
the warm absorber scenario.

Finally, we investigated model C, similar to that proposed by Piro et
al.~(2005). This produces ionization parameters consistent with zero in all
cases, thus reproducing model A.

\begin{table*}
\begin{minipage}{126mm}
\caption{1996 July observation: the best fit parameters of model B.}
\begin{tabular}{lcccccc|cccc}
\hline
s$^a$ & $\chi^2$  & N$_{H1}$$^b$ & N$_H$$w$$^c$ & $\xi$$^d$ & $\Gamma$$^e$ & $\chi^2$$^f$ &N$_{H1}$$^g$ & N$_H$$w$$^h$ & $\xi$$^i$  \\
\hline
A &  229.5/151 &  6.0$\pm_{5.3}^{3.7}$ & 20.0$\pm_{8.7}^{7.9}$ &210$\pm_{160}^{470}$ &   1.69$\pm_{0.02}^{0.02}$  & 235.1/152 &  6.3$\pm_{5.2}^{3.7}$   &  20.6$\pm_{9.0}^{7.9}$  & 220$\pm_{160}^{430}$ \\
B &  146.6/138 &   $<$5.2 & 17.6$\pm_{3.8}^{4.0}$ & 45$\pm_{35}^{170}$ &  1.72$\pm_{0.03}^{0.03}$  & 146.8/139  & $<$5.3  &  17.3$\pm_{2.3}^{4.0}$ & 40$\pm_{30}^{100}$  \\
C &  123.4/118  & $<$2.2 & 15.4$\pm_{5.3}^{5.7}$ & 140$\pm_{90}^{270}$ &  1.74$\pm_{0.07}^{0.06}$    & 124.2/119  & $<$1.8 & 14.9$\pm_{5.2}^{4.6}$   & 140$\pm_{90}^{260}$ \\
D &  139.3/118  & $<$3.3 & 13.1$\pm_{3.2}^{5.8}$ & 530$\pm_{270}^{680}$ &  1.70$\pm_{0.04}^{0.03}$  &  139.5/119  & $<$4.0 &  13.5$\pm_{2.9}^{4.9}$  & 520$\pm_{260}^{630}$ \\
\hline
\end{tabular}

\medskip

$^a$ time interval,$^b$ column density of the fully covering absorber in unit
of $10^{22}$cm$^{-2}$, $^c$ column density of uniform ionized absorber in
unit of $10^{22}$cm$^{-2}$,$^d$ absorber ionization parameter (L/nR$^2$, see
\citealt{done}),$^e$ photon index, $^f$ $\chi^2$ for $\Gamma$
fixed to 1.71, $^g$, $^h$ and  $^i$ as
$^b$, $^c$ and $^d$ respectively, for $\Gamma$ fixed to 1.71.
\end{minipage}
\end{table*}

\begin{table*}
\begin{minipage}{126mm}
\caption{ 2001 December observation: the best fit parameters of model B.}
\begin{tabular}{lcccccc|cccc}
\hline
s$^a$  & $\chi^2$  & N$_{H1}$$^b$ & N$_H$$w$$^c$ & $\xi$$^d$ & $\Gamma$$^e$ & $\chi^2$$^f$ & N$_{H1}$$^g$ & N$_H$$w$$^h$ & $\xi$$^i$ \\
\hline
J &  144.7/136&   1.5$\pm_{0.5}^{0.3}$ &  7.8$\pm_{2.1}^{2.6}$ &   360$\pm_{210}^{310}$ &   1.73$\pm_{0.03}^{0.03}$ & 146.6/137& 1.5$\pm_{0.4}^{0.3}$ & 8.3$\pm_{1.8}^{2.6}$ & 370$\pm_{90}^{330}$  \\
K &  141.9/136 &     1.7$\pm_{0.3}^{0.3}$ &  9.8$\pm_{2.1}^{2.5}$ &  460$\pm_{180}^{320}$ & 1.77$\pm_{0.02}^{0.02}$& 144.0/137& 1.8$\pm_{0.3}^{0.3}$ & 9.1$\pm_{2.0}^{2.5}$ & 440$\pm_{190}^{320}$ \\
L &  187.9/136 &      1.8$\pm_{0.3}^{0.3}$ &  9.3$\pm_{2.4}^{3.1}$ &  610$\pm_{270}^{490}$ & 1.77$\pm_{0.03}^{0.03}$&  190.4/137& 1.8$\pm_{0.3}^{0.3}$& 8.6$\pm_{2.2}^{2.9}$ & 590$\pm_{280}^{510}$ \\
M & 169.6/135&   2.0$\pm_{0.3}^{0.4}$ &   8.7$\pm_{ 2.7}^{4.7}$ & 560$\pm_{300}^{780}$ &   1.74$\pm_{0.02}^{0.02}$&  171.2/136&  2.1$\pm_{0.4}^{0.3}$& 9.4$\pm_{2.5}^{3.5}$ &  640$\pm_{320}^{680}$\\
\hline
\end{tabular}
\medskip

$^a$ time interval, $^b$ column density of the fully covering absorber in unit
of $10^{22}$cm$^{-2}$, $^c$ column density of uniform ionized absorber in
unit of $10^{22}$cm$^{-2}$,$^d$ absorber ionization parameter (L/nR$^2$, see
\citealt{done}),$^e$ photon index $^f$ $\chi^2$ for $\Gamma$
fixed to 1.76, $^g$, $^h$ and  $^i$ as
$^b$, $^c$ and $^d$ respectively, for $\Gamma$ fixed to 1.76.
\end{minipage}
\end{table*}

\section{Discussion}

We have analyzed two long (elapsed time $\gs3$ days) BeppoSAX observations of NGC
4151 searching for variability of the absorber on timescales from
tens of ksec to years . We find significant low energy spectral
variations in the softness ratio light curves of both observations suggesting that
the absorption changed during the observations.

To characterize the spectral variability we used three models as a
parameterization of the absorption: A) a partial covering absorber plus a
uniform absorber, both neutral; B) one neutral absorber and one ionized
absorber both covering the nucleus totally; and C) one ionized absorber
covering the nucleus totally and one neutral absorber covering the nucleus
only partly.  The simplest partial covering model A provided the smallest
$\chi^2$ in both 1996 July and 2001 December observations. Factor of $\sim2$
changes of both the uniform and the partial covering absorbers have been
obtained leaving the spectral index free to vary.

\begin{figure}
\begin{center}
\includegraphics[height=8truecm,width=8.5truecm]{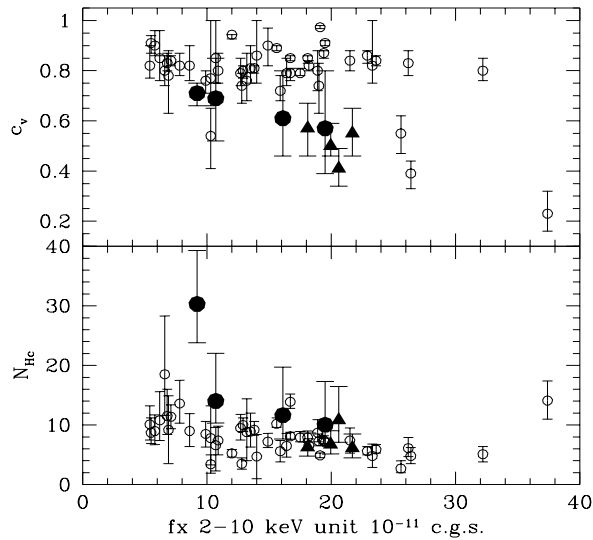}
\caption{\underline{Upper panel}: covering factor c$_v$ versus the
  2-10 keV flux. \underline{Bottom panel}: column density
N$_H$c versus 2-10 keV flux.
Solid dots identify  the best fit values for the BeppoSAX 1996 July
observation and the solid triangles identify the best fit values for the 2001
December observation. The open dots identify data
from \citealt{powa}, \citealt{yawa}, \citealt{fipe}, \citealt{yawp},
\citealt{yawa2}, \citealt{weya}, \citealt{wado}, \citealt{zdjo}.
 }
\label{nh_all}
\end{center}
\end{figure}

During the 2001 December observation the mean 6-10 keV count rate is
only slightly higher than that of the low flux intervals of the 1996
July observations, but N$_H$ and N$_H$c are a factor of about 3
smaller.  They are also smaller than those in the 1996 July C and D
intervals.  Indeed the low energy absorption during the 2001 December
observation is among the lowest ever measured in NGC 4151, as can be
seen in figure \ref{nh_all}, where we compare the best fit N$_H$c and
c$_v$ of the two BeppoSAX observations with a compilation of data from
literature.  The covering factor in the 2001 December observation
appears lower than in all other historical observations with similar
2-10 keV flux.  This suggests that on long timescales (years) the
variations of the covering factor are not (or not only) correlated
with the X-ray continuum.

The partial covering model is a convenient parameterization, but does not have
a direct physical interpretation. A more physical model is the clumpy absorber
of Holt et al. (1980). In this model the absorbing medium is composed of a
large number of small clouds, the column density of an individual cloud being
n$_c$ and the mean number of clouds along a particular line of sight being
$\mu$ \citep{homu}. In this scenario, the observed variations of the column
density are due to Poissonian fluctuations of the number of clouds, $\mu$.
The covering factor c$_v$ spans from $\sim0.4$ to $\sim0.9$ in $\sim50$
observations, with c$_v\sim 0.2$ in only one case (figure \ref{nh_all}).  An
immediate estimate of the probability of having no clouds along the line of
sight is P($\mu=0)=e^{-\mu}>$1/50, which implies $\mu>4$.  A better estimate
can be obtained assuming a gaussian distribution of c$_v$ around the average
value $<$c$_{v} >\sim 0.65$, with $\sigma(c_v)\sim0.15$, implying
$\mu\sim7$.\\ During the 1996 July observation the largest variation of the
column density is $\sim 2 \times 10^{23}$ cm$^{-2}$ on timescales as short as
$\sim 2$ days. This timescale has strong implications for the location of
the obscuring matter.  Assuming that the absorbing medium is made by spherical
clouds moving with Keplerian velocities around the central black hole
(M$_{BH}\sim 1.3\times 10^7~M_\odot$, \citealt{pefe}), and identifying the
above $\sim 2$-day timescale with the crossing time of a cloud, the
absorbing medium would be located at the distance from the nucleus, r $\leq
3.4 \times 10^4 {(n_e)}_{10}^{2} t_{2}^{2}$R$_S$, where $ {(n_e)}_{10}$ is the
density in units of 10$^{10}$ cm$^{-3}$, $t_{2}$ is the timescale in units of
two days and R$_S$ is the Schwarzschild radius \citep{riel}.  If the X-ray
absorber has a density typical of the clouds in the BELR,
10$^{9}$$<$$n_e$$<$10$^{11}$ cm$^{-3}$, then it would be located at a distance
from the nucleus 340 R$_S$$<$r$<$3.4$\times 10^6$R$_S$. This range includes
the distance of the BELR, as inferred from reverberation mapping of the
H$\beta$, C\,{\sc iv}, Mg\,{\sc ii} and H$\alpha$ line (\citealt{wape},
\citealt{clbo}, \citealt{ser}) and from theoretical estimates of the outer
radius of the BELR (\citealt{cara}). If the X-ray absorber lies in the parsec
scale dusty molecular torus (\citealt{krbe}, \citealt{pikr}, \citealt{pikr2}),
its density would be $n_e \gs5 \times 10^{10}$ cm$^{-3}$. The X-ray absorber
would then be a very compact slab (thickness $\sim 1$ R$_S$, i.e. ${\Delta
R\over{ R}}\sim 10^{-6}$) at parsec distance from the nucleus, which is a
physically unlikely structure. We then conclude that the most likely location
of the X-ray absorber is within the BELR. If the clouds absorbing the X-rays
are pressure-confining by a hot medium surrounding the BELR, this confining
gas would produce a thermal emission. Interpreting the 0.5-2 keV luminosity of
the thermal bremsstrahlung during the December 2001 observation
($\sim$10$^{41}$ \cgs), as due to this confining gas, the size of the region
containing the thermal gas is 1$\times$10$^{4}$ T$_{7}^{-1/3}$ n$_{7}^{-2/3}$
R$_S$, where T$_7$ and n$_7$ are the temperature and the density in units of
10$^7$ K and 10$^7$ cm$^{-3}$ respectively. This is consistent with the
distance from the nucleus of the BELR, and therefore a pressure-confining
medium can not be excluded.

Variations of the X-ray absorber on timescales of days are expected
in the model proposed by \citet{elvi} for type 1 AGN and extended to
type 2 AGN by \citet{riel}. In the model of \citet{elvi}, a wind
arises vertically from a narrow region on the accretion disk and it is
then radially accelerated by the radiation pressure (Murray et
al. 1995, Kartje, K{\"o}nigl \& Elitzur 1999). The wind is both warm
($\sim$10$^6$ K) and highly ionized, with a density $n_e \sim 10^{7}$
cm$^{-3}$, which puts it in pressure equilibrium with the BELR
clouds. The BELR is then a cool phase embedded in the wind and
absorbing the X-rays coming from the nucleus. In the wind model the
X-ray absorber is axisymmetric, like the torus, but the typical size
and dynamic state in the two cases are quite different. In particular
the X-ray absorber is located a few light days from the nucleus, then
changes of the column density and/or of the covering factor are
naturally expected on timescales greater than a few
days. \citet{elri} have found a factor of 100 decrease in the column
density toward the normally almost Compton-thick ($\tau \sim $0.1-0.3)
type 2 AGN NGC 4388 on timescales either $\sim$ 2 days or 4 hr, thus
giving further support to the ``wind'' model.  Such model naturally
predicts an absorber made by a cold and a warm phase.

We tested this scenario through our models B and C.  The best fit
ionization parameters are roughly in the range found by Schurch \&
Warwick (2002).  Although the behaviour of the ionization parameter in
model B is nicely consistent with the expectations, with the absorber
ionization parameter $\xi$ roughly correlated with the intrinsic flux,
the results are not satisfactory from the statistical point of view,
indicating that the physical and geometrical structure of the absorber
is more complex than any of our simple parameterizations.  Model C,
similar to that proposed by Piro et al.~(2005) produces best fit
ionization parameters consistent with zero in all cases, thus
reproducing model A. In particular, our time dependent spectral
analysis does not confirm the Piro et al. (2005) claim of a
significant highly ionized iron absorption feature. Therefore a word
of caution is necessary concernig the presence of a relativistic
outflow in this source.

To better locate the X-ray absorber, long monitoring of the source
with instruments with good energy resolution and high sensitivity (to
collect high quality spectra in a few thousands of seconds) are
needed.  Simultaneous optical monitoring of the source would allow us
to search for dust features, and therefore to understand how much dust
lies in the region occupied by the X-ray absorber.  Simultaneous X-ray
and optical-UV spectroscopy could also be used to investigate whether
broad emission lines emerge when the column density of the absorber
along the line of sight is decreasing. X-ray spectroscopy on short
timescales ($\sim 1$ ksec) could also be used to determine whether
the change of the absorbing column is step-wise (as expected for
discrete clouds with sharp edges) or more continuous, by giving a
measure of the 'fuzziness' of the clouds (i.e. their column density
gradients).

Recently, De Rosa et al. (2007) reported a time resolved spectral analysis of
the 1996 July observation, finding significant variations of the complex
absorber on timescales similar to those found in this paper. De Rosa et
al. also note, in agreement with our findings, that the level of absorbtion
during the 2001 December obervation is one of the lowest ever recorded, an
episode during which the source was unusually uncovered.

\bigskip
\centerline{\bf Acknowledgements}

This work was partially supported from NASA grant G02-3142X. We thank
Giuseppe Cesare Perola for useful discussions.


\label{lastpage}

\begin{thebibliography}{}
\bibitem[\protect\citeauthoryear{Antonucci}{1993}]{ant}Antonucci, R. 1993 ARA$\&$A, 31, 47
\bibitem[\protect\citeauthoryear{Barr et al.}{1977}]{bawh}Barr, P., White,  N. E., Sanford, P. W. 1977, MNRAS, 181, 43
\bibitem[\protect\citeauthoryear{Barvainis}{1987}]{bar}Barvainis R. 1987, ApJ, 320, 537
\bibitem[\protect\citeauthoryear{Boella et al.}{1997}]{boch}Boella, G.,  Chiappetti, L., Conti, G. et al. 1997, A$\&$AS, 122, 299
\bibitem[\protect\citeauthoryear{Burstein et al.}{1997}]{bujo}Burstein, D., Jones, C., Forman, W. et al.  1997, ApJS, 111, 163
\bibitem[\protect\citeauthoryear{Canizares}{2000}]{cani}Canizares, C. R., http://space.mit.edu/HETG/
\bibitem[\protect\citeauthoryear{Cassidy and Raine}{1997}]{cara}Cassidy I. and Raine D. J. 1997, A$\&$ A 322, 400
\bibitem[\protect\citeauthoryear{Clavel et al.}{1990}]{clbo}Clavel, J., Boksenberg, A., Bromage, G. E. et al. 1990, MNRAS, 246, 668
\bibitem[\protect\citeauthoryear{De Rosa et al.}{2007}]{deros} De Rosa, A., Piro, L.,  Perola, G. C. et al. arXiv:astro-ph/0611470
\bibitem[\protect\citeauthoryear{Done et al.}{1992}]{done}Done, C., Mulchaey, J. S., Mushotzky, R. F. et al.  1992, ApJ, 395, 275
\bibitem[\protect\citeauthoryear{Elvis et al.}{1981}]{elvis2}Elvis, M.,  Schreier, E. J., Tonry, J. et al. 1981 ApJ, 246, 20
\bibitem[\protect\citeauthoryear{Elvis, Briel $\&$ Henry,}{1983}]{elbr}Elvis, M., Briel, U. G., Henry, J. P. 1983, ApJ, 268, 105
\bibitem[\protect\citeauthoryear{Elvis et al.}{1990}]{elfa}Elvis, M., Fassnacht, C., Wilson,  A. S., Briel 1990, ApJ, 361, 459
\bibitem[\protect\citeauthoryear{Elvis}{2000}]{elvi}Elvis, M. 2000 ApJ, 545, 63
\bibitem[\protect\citeauthoryear{Elvis et al.}{2004}]{elri}Elvis M., Risaliti G., Nicastro N. et al. 2004, ApJ, 615L, 25
\bibitem[\protect\citeauthoryear{Fabbiano et al.}{1992}]{faki}Fabbiano, G., Kim, D.-W., Trinchieri, G. 1992, ApJS, 80, 531
\bibitem[\protect\citeauthoryear{Fiore et al.}{1990}]{fipe}Fiore, F., Perola, G. C., Romano, M. 1990, MNRAS, 243, 522
\bibitem[\protect\citeauthoryear{Fiore, Guainazzi $\&$ Grandi}{1999}]{figu}Fiore, F., Guainazzi, M. \& Grandi, P. 1999,Handbook for BeppoSAX NFI spectral  analysis,ftp://ftp.asdc.asi.it/pub/sax/doc/software\_docs/saxabc\_v1.2.ps.gz  or http://heasarc.gsfc.nasa.gov/docs/sax/abc/saxabc/saxabc\_v1.2.ps.gz

\bibitem[\protect\citeauthoryear{Frontera et al.}{1997}]{frco}Frontera, F., Costa, E., dal Fiume, D. et al. 1997, A$\&$AS, 122, 357
\bibitem[\protect\citeauthoryear{Gioia et al.}{1990}]{magi}Gioia, I. M., Maccacaro, T., Schild, R. E. et al. 1990, ApJS, 72, 567
\bibitem[\protect\citeauthoryear{Gursky et al.}{1971}]{guke}Gursky, H., Kellogg, E. M., Leong, C. et al. 1971, ApJ, 165L, 43
\bibitem[\protect\citeauthoryear{Holt et al.}{1980}]{homu}Holt, S. S., Mushotzky, R. F., Boldt, E. A. et al. 1980,ApJ, 241L, 13
\bibitem[\protect\citeauthoryear{Ives, Sandford \& Penston}{1976}]{ivsa}Ives, J. C. Sandford, P. W. \& Penston, M. V., 1976, AJ, 207, 159
\bibitem[\protect\citeauthoryear{Johnson et al.} {1997}]{joo} Johnson, W. N., McNaron-Brown, K., Kurfess, J. D. et al. 1997, ApJ, 482, 173
\bibitem[\protect\citeauthoryear{Kartje, K{\"o}nigl \& Elitzur}{1999}]{kako}Kartje, J. F., K{\"o}nigl, A. \& Elitzur, M. 1999, ApJ, 513, 180
\bibitem[\protect\citeauthoryear{Krolik $\&$ Begelman}{1988}]{krbe}Krolik, J. H. $\&$ Begelman M. C. 1988, ApJ, 329, 702
\bibitem[\protect\citeauthoryear{Lampton et al.}{1976}]{lama}Lampton M., Margon B. $\&$ Bowyer S., 1976, ApJ 208, 1771
\bibitem[\protect\citeauthoryear{Lawrence $\&$  Elvis}{1982}]{lael}Lawrence, A. $\&$  Elvis, M. 1982, ApJ, 256, 410
\bibitem[\protect\citeauthoryear{Magdziarz $\&$ Zdziarski}{1995}]{mazd}Magdziarz, P. $\&$ Zdziarski, A. A. 1995, MNRAS, 273, 837
\bibitem[\protect\citeauthoryear{Murphy et al.}{1996}]{mulo}Murphy, E. M.; Lockman, F. J.; Laor, A. et al. 1996 ApJS, 105, 369
\bibitem[\protect\citeauthoryear{Murray et al.}{1995}]{murr}Murray, N., Chiang, J., Grossman, S. A. et al. 1995 ApJ, 451, 498
\bibitem[\protect\citeauthoryear{}{}]{} Nicastro, F., Fiore, F., Perola, G. C., Elvis, M. 1999, ApJ, 512, 184
\bibitem[\protect\citeauthoryear{Ogle et al.}{2000}]{ogma}Ogle, P. M., Marshall, H. L., Lee, J. C. et al. 2000, ApJ, 545, 81
\bibitem[\protect\citeauthoryear{Paciesas, Mushotzky \& Pelling}{1977}]{pamu}Paciesas,  W. S., Mushotzky, R. F. \& Pelling, R. M. 1997, MNRAS, 178, 23
\bibitem[\protect\citeauthoryear{Parmar et al.}{1997}]{pama}Parmar, A. N.,  Martin, D. D. E.,  Bavdaz, M.et al. 1997, A$\&$AS, 122, 309
\bibitem[\protect\citeauthoryear{Perlman et al.}{1996}]{pest}Perlman E. S., Stocke, J. T., Wang, Q. D. et al. 1996, ApJ, 456, 451
\bibitem[\protect\citeauthoryear{Perola et al.}{1986}]{pepi}Perola, G. C., Piro, L., Altamore,  A. et al. 1986, ApJ, 306, 508
\bibitem[\protect\citeauthoryear{Peterson et al.}{2004}]{pefe}Peterson, B. M., Ferrarese, L., Gilbert, K. M. et al. 2004, ApJ, 613, 682
\bibitem[\protect\citeauthoryear{Pier $\&$ Krolik}{1992}]{pikr}Pier, E. A. $\&$ Krolik, J. H. 1992 ApJ, 399L, 23
\bibitem[\protect\citeauthoryear{Pier $\&$ Krolik}{1993}]{pikr2}Pier, E.  A. $\&$ Krolik, J. H. 1993 ApJ, 418, 673
\bibitem[\protect\citeauthoryear{Piro et al.}{2005}]{piro05}Piro L., De Rosa A., Matt G. and Perola, G. C. 2005, A\&AL, 441, L13
\bibitem[\protect\citeauthoryear{Pounds et al.}{1986}]{powa}Pounds, K. A., Warwick, R. S., Culhane,  J. L. et al. 1986, MNRAS, 218, 685
\bibitem[\protect\citeauthoryear{Risaliti, Elvis $\&$ Nicastro}{2002}]{riel}Risaliti, G., Elvis, M., Nicastro, F. 2002 ApJ, 571,  234
\bibitem[\protect\citeauthoryear{Risaliti et al.}{2005}]{risa05}Risaliti, G., Elvis, M., Fabbiano G. et al. 2005, ApJ, 623L, 93 
\bibitem[\protect\citeauthoryear{Takahashi, Inoue and Dotani}{2002}]{tain}Takahashi, K., Inoue, H. and Dotani T., 2002, PASJ, 54, 373
\bibitem[\protect\citeauthoryear{Schurch and Warwick}{2002}]{scwa}Schurch, N. J. and Warwick, R. S. 2002, MNRAS, 334, 811
\bibitem[\protect\citeauthoryear{Schurch et al.}{2003}]{scwa2}Schurch, N. J., Warwick, R. S., Griffiths R. E. et al 2003, MNRAS, 345, 423
\bibitem[\protect\citeauthoryear{Sergeev}{1994}]{ser}Sergeev, S. G. 1994,ARep, 38, 162
\bibitem[\protect\citeauthoryear{}{}]{} Smith, D. A., Georgantopoulos, I., Warwick, R. S. 2001, ApJ, 550, 635
\bibitem[\protect\citeauthoryear{Urry $\&$ Padovani}{1995}]{urpa}Urry $\&$ Padovani 1995, PASP, 107, 803
\bibitem[\protect\citeauthoryear{Yaqoob, Warwick, \& Pounds}{1989}]{yawa}Yaqoob, T., Warwick, R. S., Pounds, K. A 1989, MNRAS, 236, 153
\bibitem[\protect\citeauthoryear{Yaqoob \&  Warwick}{1991}]{yawp}Yaqoob, T. and Warwick, R. S. 1991, MNRAS, 248, 773
\bibitem[\protect\citeauthoryear{Yaqoob et al.}{1993}]{yawa2}Yaqoob, T., Warwick, R. S., Makino, F. et al.  1993, MNRAS, 264, 411
\bibitem[\protect\citeauthoryear{Yang et al.}{2001}]{yawi}Yang, Y., Wilson, A. S., Ferruit, P. et  al. 2001, ApJ, 563, 124
\bibitem[\protect\citeauthoryear{Wandel, Peterson and Malkan}{1999}]{wape}Wandel, A., Peterson, B. M. and Malkan, M. A.1999, ApJ, 526, 579
\bibitem[\protect\citeauthoryear{Warwick Done and Smith}{1995}]{wado}Warwick R. S., Done C. and Smith D. A. 1995  MNRAS, 275, 1003
\bibitem[\protect\citeauthoryear{Weaver et al.}{1994}]{weya}Weaver, K. A.,  Yaqoob,  T., Holt, S. S. et al. 1994, ApJ, 436L, 27
\bibitem[\protect\citeauthoryear{Zdziarski et al.}{1996}]{zdjo}Zdziarski, A. A., Johnson, W. N., Magdziarz, P. 1996, MNRAS, 283, 193
\bibitem[\protect\citeauthoryear{Zdziarski et al.}{2002}]{zdle}Zdziarski, A. A., Leighly, K. M., Matsuoka, M. et al. 2002, ApJ, 573, 505
\end{thebibliography}
\end{document}